\documentclass[reprint, secnumarabic,amssymb, nobibnotes, aps, prb]{revtex4-2}

\usepackage{xcolor}
\usepackage{epsfig, graphics, graphicx, subfigure, wrapfig, lipsum}
\usepackage{verbatim}
\usepackage{dcolumn}
\usepackage{bm}
\usepackage{amsmath, braket}
\usepackage{array}
\usepackage{xr}
\usepackage{multirow}
\usepackage{longtable}
\usepackage{tabularray}

\newcolumntype{X}[1]{>{\centering\arraybackslash\hspace{0pt}}p{#1}}
\newcolumntype{M}[1]{ >{\centering\arraybackslash}m{#1}}
\newcommand{\roml}[1]{\lowercase\expandafter{\romannumeral #1\relax}}
\newcommand{\romu}[1]{\uppercase\expandafter{\romannumeral #1\relax}}


\begin{document}

\title{Computational uncertainties in lattice thermal conductivity prediction of crystalline solids}

\author{Yagyank Srivastava}
\author{Amey G. Gokhale}
\author{Ankit Jain}
\email{a\_jain@iitb.ac.in}
\affiliation{Mechanical Engineering Department, IIT Bombay, India}
\date{\today}%

\begin{abstract}
We report computational uncertainties in Boltzmann Transport Equation (BTE)-based lattice thermal conductivity prediction of 50 diverse semiconductors from the use of different BTE solvers (ShengBTE, Phono3Py, and in-house code) and interatomic forces. The interatomic forces are obtained either using the density functional theory (DFT) as implemented in packages Quantum Espresso and VASP employing commonly used exchange correlation functionals (PBE, LDA, PBEsol, and rSCAN) or using the pre-trained foundational machine learning forcefields trained on two different material datasets. 

We find that the considered BTE solvers introduce minimal uncertainties and, using the same interatomic force constants, all solvers result in an excellent agreement with each other, with a mean absolute percentage error (MAPE) of only 1\%. While this error increases to around 10\% with the use of different DFT packages, the error is still small and can be reduced further with the use of stringent planewave energy cutoffs. On the other hand, the differences in thermal conductivity due to the use of different exchange correlation functionals are large, with a MAPE of more than 20\%. 
The currently available pre-trained foundational ML models predict the right trend for thermal conductivity, but the associated errors are high, limiting their applications for coarse screening of materials.

\end{abstract}
\maketitle

\section{Introduction}
The thermal conductivity of a material plays a crucial role in determining its performance in various technological applications.\cite{clarke2003,dames2005,lindsay2018}.  
In semiconducting and insulating solids, the thermal transport is primarily due to atomic vibrations, i.e., phonons \cite{reissland1973, dove1993, wallace1972}. Conventionally, the search of materials with low- and high-thermal conductivity is led by experiments and/or empirical experimental observations \cite{slack1973}. However, in the last decade, with the availability of computational resources and computer codes, it has become possible to predict phonon thermal conductivity from ab initio by solving the Boltzmann transport equation (BTE) \cite{ward2008, esfarjani2011,  lindsay2018, mcgaughey2019, jain2025}. 

The calculation of phonon thermal conductivity ($\kappa$) based on the Boltzmann transport equation (BTE) approach begins with the evaluation of interatomic forces \cite{mcgaughey2019, jain2025}.
Traditionally, these forces were obtained from density functional theory (DFT) calculations by employing different exchange-correlation (XC) functionals \cite{jain2015b}. With advances in machine learning models in the past few years, it is now possible to get these forces from pre-trained [on DFT forces] foundational machine learning (ML) models \cite{pota2025}.  The interatomic forces obtained from DFT or ML models are then employed to extract interatomic force constants (using tools such as thirdorder.py \cite{shengBTE2014}, hiPhive \cite{hiphive2019}, etc) and the obtained interatomic force constants are subsequently passed on to BTE solvers (such as ShengBTE \cite{shengBTE2014}, Phono3py \cite{phono3py}, etc) to get to the $\kappa$ of material. In addition to the employed thermal transport theory (such as three- vs four-phonon scattering, phonon renormalization, etc), the accuracy of thermal conductivity obtained using this BTE-based approach depends on the interatomic forces and employed BTE solvers. There are multiple reports in the literature discussing the role of the employed thermal transport theory on predicted $\kappa$ for various material systems \cite{xia2020, jain2020, jain2022}. The studies on the effect of force interactions (via employed XC functional/ML-model  DFT package) and the employed BTE package are rare, limited primarily to a handful of simple cubic material systems \cite{jain2015b, xia2020b}. Nevertheless, for the computational discovery of novel material systems, it is important to establish the uncertainties in $\kappa$ arising from the choice of these simulation parameters.

In this work, we carried out a computational study to systematically quantify the uncertainties in predicting $\kappa$ originating from employed interatomic forces (based on different DFT solvers, XC functionals, ML methods) and from employed BTE solvers, on 50 diverse semiconducting materials. We employed commonly used DFT packages Quantum Espresso (QE) \cite{giannozzi2009} and VASP \cite{vasp}, both based on planewave basis-set, and compared LDA \cite{kohn1965}, PBE \cite{perdew1996}, PBEsol \cite{perdew2008}, and rSCAN \cite{bartok2019} functionals. We also compared the performance of two variants of the foundational ML model MACE \cite{mace} trained on the MaterialsProject (MP) \cite{materialsProject} and MatPES datasets \cite{matpes}. The $\kappa$ calculations via the BTE solution are performed using our in-house code (ALD), ShengBTE, and Phono3py, to establish uncertainties from BTE solvers.

We find that there are minimal differences in $\kappa$ obtained from different BTE solvers. The maximum variation in $\kappa$ is due to the choice of the employed functional. 
{\color{black}Surprisingly, we find that the commonly employed PBE functional results in an under-prediction of $\kappa$ compared to other functionals (17\% under-prediction compared to PBEsol functional). This is alarming as many of the large material datasets are constructed using PBE functional \cite{materialsProject, matpes}. As such, ML forcefields trained on such datasets may result in a similar under-prediction of $\kappa$. Further, a quick search on Google Scholar with keywords ``PBE functional'' and ``phonon thermal conductivity'' returns 264 hits compared to only 42 and 47 for LDA and PBEsol functionals, indicating wider use of the PBE functional for $\kappa$ calculations.} 
Finally, the errors obtained using foundational ML models are currently large, but these models are able to get the correct trend of $\kappa$.

\section{Methodology}
\label{sec_method}
We calculate the thermal conductivity, $k_{\alpha}$, in the $\alpha$-direction by solving the BTE and using Fourier's law as \cite{reissland1973, jain2020}:
\begin{equation}
 \label{eqn_theory_conduct}
 k_{\alpha} = \sum_{{\boldsymbol{q}}} \sum_{\nu} c_{{\boldsymbol{q}}\nu} v_{{\boldsymbol{q}}\nu, \alpha}^{2} \tau_{{\boldsymbol{q}}\nu, \alpha}.
\end{equation}
The summation in the Eqn.~\ref{eqn_theory_conduct} is over all the phonon wavevectors, ${\boldsymbol{q}}$, and polarizations, $\nu$ and  $c_{{\boldsymbol{q}}\nu}$ is the phonon specific heat, $v_{{\boldsymbol{q}}\nu,\alpha}$ is the $\alpha$ component of phonon group velocity vector ${\boldsymbol{v}_{{\boldsymbol{q}}\nu}}$, and $\tau_{{\boldsymbol{q}}\nu, \alpha}$ is the phonon scattering time. Phonons are bosons and follow the Bose-Einstein distribution, when in equilibrium. The phonon specific heat can be obtained from the phonon vibrational frequencies as:
\begin{equation}
\label{eqn_theory_cph}
c_{{\boldsymbol{q}}\nu} = \frac{\hbar\omega_{{\boldsymbol{q}}\nu}}{V} \frac{\partial n^{o}_{{\boldsymbol{q}}\nu}}{\partial T} = \frac{k_\text{B} x^2 e^x }{(e^x-1)^2}.
\end{equation}
The $n^o_{{\boldsymbol{q}}\nu}$ in Eqn.~\ref{eqn_theory_cph} is the Bose-Einstein distribution ($n^o_{{\boldsymbol{q}}\nu} = \frac{1}{e^x-1}$), $\hbar$ is the reduced Planck constant, $\omega_{{\boldsymbol{q}}\nu}$ is the phonon frequency, $V$ is the crystal volume,  $T$ is the temperature, $k_\text{B}$ is the Boltzmann constant, and $x = \frac{\hbar\omega_{{\boldsymbol{q}}\nu}}{k_\text{B}T}$. The phonon group velocities are obtained as:
\begin{equation}
    {\boldsymbol{v}_{{\boldsymbol{q}}\nu}} = \frac{\partial \omega_{\boldsymbol{q}\nu}}{\partial \boldsymbol{q}}.
\end{equation}
The phonon frequencies are obtained from the diagonalization of the dynamical matrix as:
\begin{equation}
 \label{eqn_theory_evproblem}
 \omega_{{\boldsymbol{q}}\nu}^{2} {\boldsymbol{e}}_{{\boldsymbol{q}}\nu} = {\boldsymbol{D}}_{{\boldsymbol{q}}} \cdot {\boldsymbol{e}}_{{\boldsymbol{q}}\nu},
\end{equation}
where ${\boldsymbol{e}}_{{\boldsymbol{q}}\nu}$ is the eigenvector and ${\boldsymbol{D}}_{{\boldsymbol{q}}}$ is the Dynamical matrix  whose elements, $D_{\boldsymbol{q}}^{3(b-1)+\alpha, 3(b^{'}-1) + \beta}$, given by \cite{dove1993}:
\begin{equation}
 \label{eqn_theory_dynamical}
 \begin{split}
 D_{\boldsymbol{q}}^{3(b-1)+\alpha, 3(b^{'}-1) + \beta} = \frac{1}{\sqrt{m_b m_{b^{'}}}} \sum_{l^{'}} \Phi_{b0;b^{'}l^{'}}^{\alpha\beta} 
 \\
 \exp{\{i[{\boldsymbol{q}}.( {\boldsymbol{r}}_{b^{'}l^{'}}  -  {\boldsymbol{r}}_{b0}  )] \}},
 \end{split}
\end{equation}
where the summation is over all unit-cells in the lattice. Here $m_b$ is the mass of atom $b$ in the unit-cell,  $\boldsymbol{r}_{bl}$ is the position vector of atom $b$ in the $l^{th}$ unit-cell, and $\Phi_{ij}^{\alpha\beta}$ is the real-space ($ij, \alpha\beta$)-element of the  harmonic force constant matrix $\boldsymbol{\Phi}$.
The phonon scattering time is obtained by considering the three-phonon scattering processes as \cite {reissland1973, wallace1972, jain2020}:

\begin{equation}
\begin{split}
 \label{eqn_rta_3ph}
 \frac{1}{\tau_{\boldsymbol{q}\nu}^{}} 
 =
  \sum_{{\boldsymbol{q}_{1}}\nu_{1}}
 \sum_{{\boldsymbol{q}_{2}}\nu_{2}}
 \bigg\{
 \Big\{
  {(n_{{\boldsymbol{q}_{1}\nu_{1}}}  - n_{{\boldsymbol{q}_{2}\nu_{2}}})}
 W^{+}
  \Big\} 
  +
  \\
  \frac{1}{2}
   \Big\{
  (n_{{\boldsymbol{q}_{1}\nu_{1}}} + n_{{\boldsymbol{q}_{2}\nu_{2}}} +  1)
  W^{-}
   \Big\}
   \bigg\},
 \end{split}
 \end{equation}
where $\boldsymbol{W}$ represents scattering probability matrix given by:
\begin{equation}
    \begin{split}
    \label{eqn_W_3ph}
W^{\pm}       
=
\frac{2\pi}{\hbar^2}
 \left|
 \Psi_{ {\boldsymbol{q}} (\pm{\boldsymbol{q}_{1}}) (-{\boldsymbol{q}_{2}}) }^{\nu \nu_{1} \nu_{2}}
 \right|^2
  \delta({\omega_{{{\boldsymbol{q}}_{}}\nu_{}} \pm \omega_{{{\boldsymbol{q}}_{1}}\nu_{1}} - \omega_{{{\boldsymbol{q}}_{2}}\nu_{2}}}).
    \end{split}
\end{equation}
The $\Psi_{ {\boldsymbol{q}} {\boldsymbol{q}_{1}} {\boldsymbol{q}_{2}} }^{\nu \nu_{1} \nu_{2}}$ are the Fourier transform of real-space cubic constants, $\Psi^{\alpha\beta\gamma}_{bl;b^{'} l^{'};b^{''} l^{''}}$, and are obtained as:
\begin{equation}
    \begin{split}
        \label{eqn_cubic_IFC}
\Psi_{ {\boldsymbol{q}} {\boldsymbol{q}_{1}} {\boldsymbol{q}_{2}} }^{\nu \nu_{1} \nu_{2}}
=
 \Psi_{ {\boldsymbol{q}} {\boldsymbol{q}^{'}} {\boldsymbol{q}^{''}} }^{\nu \nu^{'} \nu^{''}} = 
 N
 {\left(\frac{\hbar}{2N}\right)}^{\frac{3}{2}}
 \sum_{b} \sum_{b^{'} l^{'}}
\sum_{b^{''} l^{''}} 
\sum_{\alpha\beta\gamma} 
\Psi^{\alpha\beta\gamma}_{bl;b^{'} l^{'};b^{''} l^{''}}
\\
\times
\frac{
{{\tilde{e}}_{b,\boldsymbol{q}\nu}^{\alpha}}  
{{\tilde{e}}_{b^{'},{\boldsymbol{q}}^{'} \nu^{'}}^{\beta}} 
{{\tilde{e}}_{b^{''},{\boldsymbol{q}}^{''} \nu^{''}}^{\gamma}} }{\sqrt{ 
{m_b \omega_{\boldsymbol{q}\nu}}  
{m_{b^{'}} \omega_{{\boldsymbol{q}}^{'}\nu^{'}}}   
{m_{b^{''}} \omega_{{\boldsymbol{q}}^{''}\nu^{''}}}   }}  
e^{[i( {{\boldsymbol{q}}^{'}}  \cdot{\boldsymbol{r}}_{0l^{'}} 
+    {{\boldsymbol{q}}^{''}}  \cdot{\boldsymbol{r}}_{0l^{''}} )]} ,
    \end{split}
\end{equation}
The delta-function, $\delta$ in Eqn.~\ref{eqn_W_3ph} ensures energy conservation. The summation in Eqn.~\ref{eqn_cubic_IFC} is performed over phonon wavevectors satisfying crystal momentum conservation, i.e., $\boldsymbol{q} + \boldsymbol{q_1} + \boldsymbol{q_2} = \boldsymbol{G}$, where $\boldsymbol{G}$ is the reciprocal space lattice vector.

\subsection{Computational Details}
The computation of phonon thermal conductivity via Eqn.~\ref{eqn_theory_conduct} requires harmonic and anharmonic interatomic force constants. The harmonic force constants are obtained using a four-point central difference scheme with a perturbation size of $0.03$ $\text{\AA}$ on conventional unit-cell-based supercells. The electronic Brillouin zone is sampled using Monkhorst-Pack wavevector grid of size $N^k_i$ such that $N^k_i.|a_i^{comp}| \sim 30$ $\text{\AA}$, where $|a_i^{comp}|$ represents the length of computational cell lattice vector $\textbf{a}_i^{comp}$. For Quantum Espresso \cite{giannozzi2009}, the calculations are performed using sg15 Optimized Norm-conserving Vanderbilt pseudopotentials \cite{hamann2013} and the planewave kinetic energy cutoff is set at 80 Ry in all calculations.  The electronic total energy is converged to within $10^{-10}$ Ry/atom during self-consistent cycles and the structure relaxations are performed with a force convergence criterion of $10^{-5}$ Ry/$\text{\AA}$. For VASP, the calculations are performed using LDA, PBE, PBEsol, and rSCAN functionals with the planewave energy cutoffs of $1.3\times$ENMAX, where ENMAX is the maximum ENCUT reported in the pseudopotential files of participating species. The electronic total energy is converged to within $10^{-6}$ eV during self-consistent cycles and the structure relaxations are performed with an energy convergence criterion of $10^{-4}$ eV. 

The harmonic force constants are obtained from perturbed supercells generated using $N^{har}_i$ repetitions of the conventional unitcell such that $N^{har}_i.|a_i^{conv}| \sim 20$ $\text{\AA}$ using Gamma-point sampling of the electronic Brillouin zone. The cubic and quartic IFCs are obtained from Taylor-series fitting of Hellmann-Feynman forces obtained on 200 thermally populated supercells  (corresponding to a temperature of 300 K) obtained from $N^{anhar}_i$ repetitions of the conventional unitcell such that $N^{anhar}_i.|a_i^{conv}| \sim 15$ $\text{\AA}$, where $|a_i^{conv}|$ represents the length of the conventional unitcell lattice vector $\textbf{a}_i^{conv}$.  The cubic and quartic force constant interaction cutoffs are set at $6.5$ and $4.0$ $\text{\AA}$ for the majority of compounds, though for some compounds with lower symmetries, these values are reduced to $5.0$ and $2.5$ $\text{\AA}$. For the computation of three-phonon scattering rates, the phonon wavevector grid of size $N^q_i$ is employed such that $N^q_i.|a_i^{comp}| \sim 100$ $\text{\AA}$.

We note that since the primary objective of this work is to do benchmarking of forces/BTE solver, all reported thermal conductivities are obtained using the relaxation time approximation of the BTE, with only three-phonon scattering, and by considering only the particle-channel contribution to thermal transport.

\subsection{Materials Selection}
We selected ternary materials from the Materials Project \cite{materialsProject} based on the following criterion: (\roml{1}) removed materials belonging to triclinic,  monoclinic, and orthorhombic spacegroups (\roml{2}) removed materials containing lanthanides and actinides, noble gases, and precious metals, (\roml{3}) removed strongly ionic compounds formed by halides, oxides, and hydrides, (\roml{4}) removed materials with electronic bandgap lower than $0.2$ eV,  (\roml{5}) limited to thermodynamically stable materials with energy above the convex hull (with respect to all reported materials in the Materials Project) less than $0.2$ eV/atom, and (\roml{6}) removed materials with more than 15 atoms in the unitcell. These filters resulted in 429 materials,  of which the full thermal conductivity calculations are successful (without any imaginary phonon modes from any of the considered functional/DFT packages) on 48 materials. The dataset also includes one binary and one quaternary material. 

\subsection{MAPE and average properties}
We define the mean absolute percentage error (MAPE) on the basis of average values as $MAPE = 2|\frac{y-x}{y+x}|\times100\%$, where $y$, $x$ are values obtained from two different functionals, BTE solvers, or DFT packages. Notice that $(x+y)/2$ in this definition corresponds to the average value. Similarly, the mean percentage error is also defined on the basis of average values as $MPE = 2\frac{y-x}{y+x}\times100\%$.

The heat capacity weighted average 
Gr\"uneissen parameters are obtained as 
$\frac{\sum_i c_i |\gamma_i|}{\sum_i c_i}$, where the summations are over all phonon modes and $c_i$, 
and $\gamma_i$ are the mode-dependent heat capacity, 
and Gr\"uneissen parameter.

\section{Results}
\begin{figure}
\begin{center}
\epsfbox{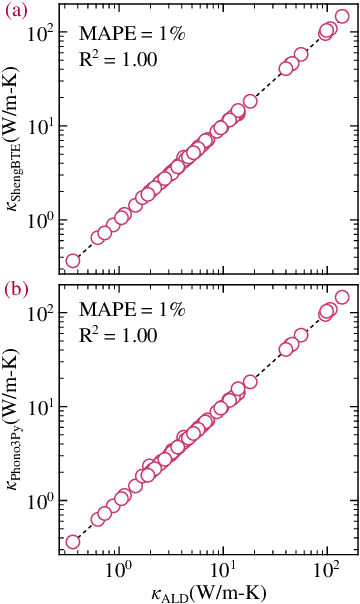}
\end{center}
\caption{The effect of the employed BTE solver on the predicted thermal conductivity. The comparison is made between (a) ShengBTE and ALD (in-house) BTE solvers, and (b) Phono3Py and ALD BTE solvers. All results are obtained using the same set of interatomic force constants (by converting between file formats).   The interatomic forces are obtained by employing the PBE functional with the QE DFT package. The results suggest less than 1\% variation in $\kappa$ with different solvers.}
\label{fig_BTE}
\end{figure}

{\bf BTE Solver:} In Fig.~\ref{fig_BTE}, we first report the effect of the employed BTE solver on the obtained $\kappa$. For this, we obtain interatomic forces using the QE DFT package \cite{giannozzi2009} with PBE XC functional \cite{perdew1996}. We computed harmonic force constants using finite-difference with a step size of $0.03$ $\text{\AA}$. The anharmonic force constants are obtained using the thermal snapshot method with 200 snapshots \cite{jain2020}. To remove dependence on interatomic force constants, we used the same set of interatomic force constants with different solvers along with the same thermal transport physics (three-phonon and isotope scatterings). 

In Fig.~\ref{fig_BTE}(a), the $\kappa$ obtained using the ShengBTE solver \cite{shengBTE2014} are compared with those obtained using the ALD solver, and in Fig.~\ref{fig_BTE}(b), the $\kappa$ obtained using the Phono3py solver \cite{phono3py} are compared with those obtained using the ALD solver. We find that the $\kappa$ obtained using various BTE  solvers shows minimal differences, and the obtained coefficient of determination, $\text{R}^2$, is perfect at  $1.0$ for all solvers.  This agreement is really impressive as the employed solvers have different numerical implementations and handle energy conservation delta functions using different approaches (tetrahedron, Gaussian, and Lorentzian in Phono3py, ShengBTE, and ALD). The maximum obtained difference between $\kappa_\text{avg,BTE} = 1/3\times(\kappa_\text{ALD} + \kappa_\text{ShengBTE}  + \kappa_\text{Phono3py})$ and $\kappa_\text{ALD}$, $\kappa_\text{ShengBTE}$, $\kappa_\text{Phono3py}$ is 7\%, 5\%, 12\% for $\text{CdLiAs}$ (cubic), $\text{SrIn}_2\text{As}_2$ (c-axis, hexagonal), $\text{SrIn}_2\text{As}_2$ (c-axis, hexagonal) respectively.

\begin{figure}
\begin{center}
\epsfbox{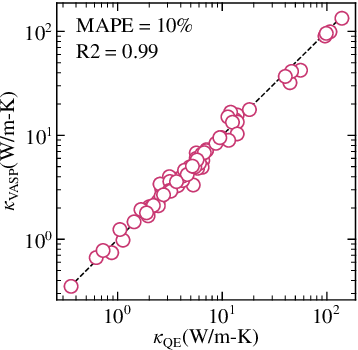}
\end{center}
\caption{The effect of using different DFT packages for obtaining interatomic forces for the thermal conductivity calculations. The forces are obtained using the PBE functional via VASP and QE DFT packages by using planewave energy cutoff of 1.3$\times$ENMAX and 80 Ry respectively.}
\label{fig_DFT}
\end{figure}

{\bf DFT Package:} In Fig.~\ref{fig_DFT}, we compare $\kappa$ obtained from VASP-based DFT forces against those obtained using QE-based DFT forces. We employed the PBE XC functional in both packages and obtained $\kappa$ via the ALD BTE solver.  All calculations are performed using exactly the same simulation parameters (barring planewave energy cutoff and pseudopotentials). 

We find that the $\kappa$ obtained from the forces of these two DFT packages follows a similar trend and has a high $\text{R}^2$ of $0.99$ with MAPE 10\%. Some variations in the results from these considered DFT packages are expected due to different pseudo-potentials and different planewave energy cutoffs: the planewave energy cutoff in QE is fixed at 80 Ry while in VASP, it is set at 1.3$\times$ENMAX. 

The maximum obtained difference between $\kappa$ from two DFT packages is for $\text{CuGa}_5\text{Se}_8$ in the tetragonal spacegroup for which $\kappa$ obtained using QE package-based forces is $6.05$ and 5.28 W/m-K along two unequal directions, while that obtained using the VASP-based forces is 5.38 and 3.32 W/m-K respectively.  However, this large difference is due to the employed planewave energy cutoff and with the use of 120 Ry cutoff in QE-based force evaluations, the $\kappa$ obtained from QE changes to 5.54 and 3.61 W/m-K (within 9\% of those obtained using the VASP-based forces). We expect a similar agreement for other materials with changes in the planewave energy cutoff. However, since our employed values of planewave energy cutoffs are typical for the considered DFT packages, we decided not to refine calculations any further and reported differences originating from these typical values.

{\bf XC Functional:} 

\begin{figure*}
\begin{center}
\epsfbox{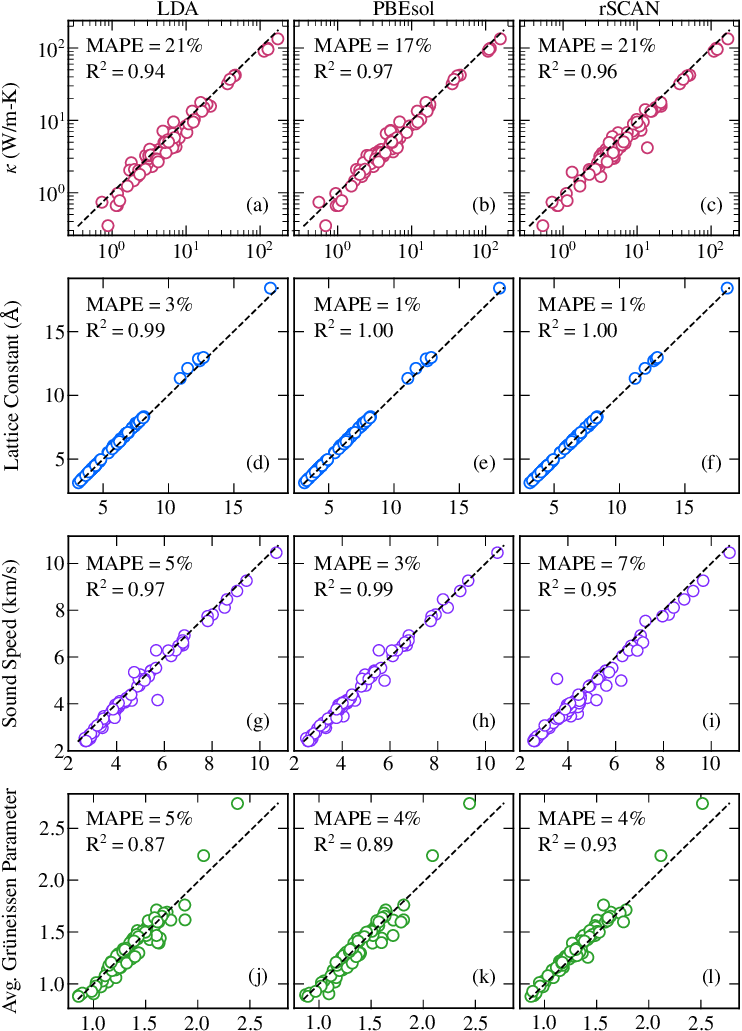}
\end{center}
\caption{The effect of exchange correlation functional on the thermal transport properties. The properties reported on the y-axis are all obtained using the PBE functional, while those reported on the x-axis are obtained using LDA, PBEsol, and rSCAN functionals for left, center, and right columns. All calculations are performed using the VASP DFT package and $\kappa$ are obtained using the ALD BTE solver. The Gr\"uneissen parameters are heat capacity weighted.}
\label{fig_XC}
\end{figure*}

We next investigate the variation in $\kappa$ due to the use of different XC functionals in DFT calculations. For this, we obtained forces from the VASP DFT package and solved BTE using the ALD solver. We employed commonly used PBE \cite{perdew1996}, LDA \cite{kohn1965},  PBEsol \cite{perdew2008}, and rSCAN \cite{bartok2019} functionals. {\color{black} These functionals differ in their treatment of electron density, gradient corrections, and kinetic energy density, which influences their accuracy in predicting structural and energetic properties. While LDA tends to over-bind and PBE typically overestimates lattice constants, PBEsol improves structural predictions for solids, and rSCAN (numerically stable revision of SCAN), as a meta-GGA, provides enhanced accuracy across diverse bonding environments.} 

The results obtained using the PBE functional are compared with LDA,  PBEsol, and rSCAN functionals in Figs.~\ref{fig_XC}(a), \ref{fig_XC}(b), and \ref{fig_XC}(c), where the values reported on the ordinate axis are obtained using the PBE functional, while those on the abscissa are obtained using LDA, PBEsol, and rSCAN functionals for the left, center, and right columns.

We find that the employed functional has a profound effect on the obtained $\kappa$, and the MAPE between different functionals varies to over 20\%. For the majority of the materials, we find that $\kappa$ obtained using the PBE {(\color{black}rSCAN)} functional-based forces is lower {\color{black}(higher)} than that obtained using other functionals. To understand the origin of this large variation in $\kappa$ from different functionals, we report lattice constant, speed of sound, and heat-capacity weighted Gr\"uneissen parameters in Figs.\ref{fig_XC}(d)-\ref{fig_XC}(f), \ref{fig_XC}(g)-\ref{fig_XC}(i), and \ref{fig_XC}(j)-\ref{fig_XC}(l) respectively. {\color{black}Further, the average performances of each of these considered functionals are also reported in Table ~\ref{table_XC} for $\kappa$, lattice constants, speed of sound, and average Gr\"uneissen parameter. } {\color{black} The sound speeds are obtained as group velocities of longitudinal acoustic phonons in the long wavelength limit and are indicative of the correctness of harmonic interatomic force constants. Similarly, Gr\"uneissen parameters are a measure of the anharmonicity of material and are indicative of the correctness of cubic interatomic force constants. }

\begin{table*}[]
\caption{The correlation between thermal transport properties obtained from different exchange correlation functionals. The average values are obtained by taking the average of all functionals for a given material. The reported numbers in {\color{violet}violet} are the mean absolute percentage error obtained by taking the mean of absolute percentage errors, $\frac{|x-y|}{(x+y)/2}\times 100$ over considered materials. Similarly, the numbers in {\color{black}black} are the mean percentage error obtained by taking the mean of percentage errors, $\frac{(x-y)}{(x+y)/2}\times 100$ over considered materials. $x$, $y$ denote values from rows and columns respectively.}

\begin{tabular}{|c|M{1.8cm}||M{1.2cm}|M{1.2cm}|M{1.2cm}|M{1.2cm}||M{1.8cm}|}
\hline

& & PBE & LDA & PBEsol & rSCAN & Average \\
\hline

 & PBE & - & {\color{violet}21\%} & {\color{violet}17\%} & {\color{violet}21\%} & {\color{violet}14\%} \\
Thermal & LDA & {\color{black} 15\% } & - & {\color{violet}10\%} & {\color{violet}18\%} & {\color{violet}10\%} \\
Conductivity & PBEsol & {\color{black} 9\% } & {\color{black} -7\% } & - & {\color{violet}19\%} & {\color{violet}8\%} \\
& rSCAN & {\color{black} 18\% } & {\color{black} 2\% } & {\color{black} 9\% } & - & {\color{violet}12\%} \\
& Average & {\color{black} 12\% } & {\color{black} -4\% } & {\color{black} 3\% } & {\color{black} -6\% } & - \\
\hline

 & PBE & - & {\color{violet}3\%} & {\color{violet}1\%} & {\color{violet}1\%} & {\color{violet}1\%} \\
Lattice & LDA & {\color{black} -3\% } & - & {\color{violet}1\%} & {\color{violet}2\%} & {\color{violet}1\%} \\
Constant & PBEsol & {\color{black} -1\% } & {\color{black} 1\% } & - & {\color{violet}1\% }& {\color{violet}0\%} \\
& rSCAN & {\color{black} -1\% } & {\color{black} 2\% } & {\color{black} 1\% } & - & {\color{violet}0\%} \\
& Average & {\color{black} -1\% } & {\color{black} 1\% } & {\color{black} 0\% } & {\color{black} 0\% } & - \\
\hline

 & PBE & - & {\color{violet}5\%} & {\color{violet}3\%} & {\color{violet}7\%} & {\color{violet}4\%} \\
Speed & LDA & {\color{black} 5\% } & - & {\color{violet}3\%} & {\color{violet}5\%} & {\color{violet}3\%} \\
of & PBEsol & {\color{black} 2\% } & {\color{black} -2\% } & - & {\color{violet}5\%} & {\color{violet}2\%} \\
Sound & rSCAN & {\color{black} 6\% }& {\color{black} 1\% } & {\color{black} 3\% } & - & {\color{violet}4\%} \\
& Average & {\color{black} 3\% } & {\color{black} -1\% } & {\color{black} 1\% } & {\color{black} -2\% } & - \\
\hline

& PBE & - & {\color{violet}5\%} & {\color{violet}4\%} & {\color{violet}4\%} & {\color{violet}2\%} \\
Average & LDA & {\color{black} 1\% } & - & {\color{violet}2\%} & {\color{violet}6\%} & {\color{violet}3\%} \\
Gr\"uneissen & PBEsol & {\color{black} 1\% } & {\color{black} 0\% } & - & {\color{violet}6\%} & {\color{violet}2\%} \\
Parameter & rSCAN & {\color{black} -3\% } & {\color{black} -4\% } & {\color{black} -4\% } & - & {\color{violet}4\%} \\
& Average & {\color{black} 0\% } &{\color{black} -1\% } & {\color{black} -1\% } & {\color{black} 3\% } & - \\
\hline

\end{tabular}
\label{table_XC}
\end{table*}

{\color{black} As reported in literature, we find that lattice constant trends are systematic across the considered materials with $a_{\text{LDA}} < a_{\text{PBEsol}} < a_{\text{rSCAN}} < a_{\text{PBE}} $, suggesting over- and under-binding from LDA and PBE functionals. This over- and under-binding by LDA and PBE functionals results in stiffened and softened phonon dispersions from two functionals. This is reflected in the speed of sound, which is consistently under-/over-predicted from the PBE/LDA functionals compared to the PBEsol functional, with a mean absolute percentage error of 3\% each. Due to this, the $\kappa$ obtained from the PBE/LDA functional are majorly under-/over-predicted compared to the PBEsol functional. However, the mean absolute percentage error in predicting $\kappa$ using the PBE functional is 17\% compared to PBEsol, whereas the LDA functional yields more consistent results with PBEsol, with a corresponding error of 10\%.}

\begin{figure}
\begin{center}
\epsfbox{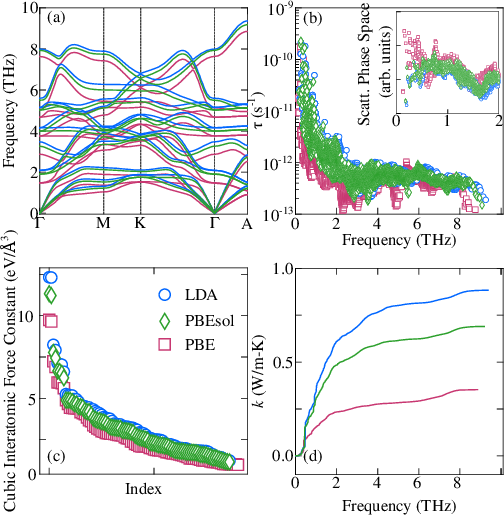}
\end{center}
\caption{{\color{black}Phonon thermal transport properties of $\text{AgScSe}_2$ (trigonal space group) calculated using PBE, PBEsol, and LDA functionals: (a) phonon dispersion relations, (b) phonon scattering lifetimes with an inset showing the three-phonon scattering phase space for low-frequency modes, (c) the most significant (largest magnitude) cubic interatomic force constants, and (d) thermal conductivity accumulation as a function of phonon frequency. The accumulation reported in (d) is for cross-plane thermal transport.}}
\label{fig_AgScSe}
\end{figure}

{\color{black} To understand this large difference from the PBE functional, we focus on $\text{AgScSe}_2$ in the trigonal spacegroup, for which the $\kappa$ is consistently under-predicted with respect to all other functionals, with an obtained value of 0.35  W/m-K from the PBE functional compared to 0.88, 0.69, and 0.53 W/m-K from LDA, PBEsol, and rSCAN functionals in the cross-plane direction [Figs.~\ref{fig_XC}(a)-\ref{fig_XC}(c)]. The mode-dependent phonon properties of $\text{AgScSe}_2$ are reported in Fig.~\ref{fig_AgScSe}. As expected from the lattice constant, the phonon dispersions obtained using the PBE/LDA functionals are soft/stiff compared to the PBEsol functional [Fig.~\ref{fig_AgScSe}(a)]. However, despite lower values of cubic interatomic force constants [Fig.~\ref{fig_AgScSe}(c)], the obtained phonon scattering lifetimes are also lower from the PBE functional compared to the other two functionals. This is an indirect effect of phonon softening, which results in acoustic-bunching, making it easier for phonons to satisfy momentum and energy conservation selection rules. This is reported in the inset of Fig.~\ref{fig_AgScSe}(b) for low-frequency acoustic phonons, where the obtained three-phonon scattering phase space from the PBE functional is larger than that from the other functionals. Consequently, the lower phonon group velocities combined with reduced phonon lifetimes result in a less than 0.25 W/m-K contribution from low-frequency acoustic phonons (frequency lower than 2 THz) from the PBE functional compared to more than 0.5 W/m-K from other functionals [Fig.~\ref{fig_AgScSe}(d)].}

{\color{black}For rSCAN functional, the obtained lattice constants are similar to those from PBE and PBEsol functionals (mean absolute error of only 1\%), and the obtained phonon group velocities are over-predicted compared to all other functionals (with the exception of  LiTaS for which rSCAN functional velocities are lower than all other functionals \cite{Ta_note}). Further, the rSCAN functional results in an under-prediction of lattice anharmonicity compared to all other functionals as reflected in average Gr\"uneissen parameters (Table~\ref{table_XC}).  Consequently, the $\kappa$ obtained from the rSCAN functional are higher, on average, compared to those from other functionals. The largest deviation of $\kappa$ obtained using the rSCAN functional is for ZnLiAs in the cubic space group with a value of 13.66 W/m-K compared to 4.19, 6.63, and 6.38 W/m-K from PBE, LDA, and PBEsol functionals.}

\begin{figure}
\begin{center}
\epsfbox{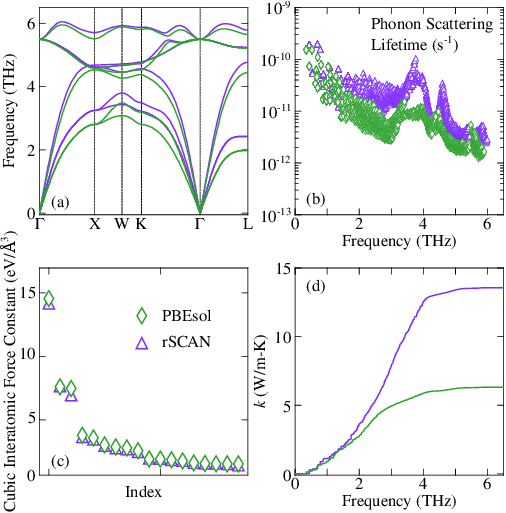}
\end{center}
\caption{{\color{black}Phonon thermal transport properties of $\text{ZnLiAs}$ (cubic space group) calculated using PBEsol and rSCAN functionals: (a) phonon dispersion relations, (b) phonon scattering lifetimes, (c) the most significant (largest magnitude) cubic interatomic force constants, and (d) thermal conductivity accumulation as a function of phonon frequency. The $\kappa$ obtained using the rSCAN functional is a factor of two larger than that from the PBEsol functional owing to stiffened phonon dispersion (over-prediction of phonon group velocities) and reduced acoustic bunching (reduced phonon scattering phase space and larger phonon scattering lifetimes).}}
\label{fig_ZnLiAs}
\end{figure}

{\color{black} The mode-dependent phonon properties of ZnLiAs as obtained from the rSCAN functional are compared with those from the PBEsol functional in Fig.~\ref{fig_ZnLiAs}. The speed of sound obtained from rSCAN is more than 5500 m/s compared to less than 5000 m/s from the PBEsol functional. This stiffened acoustic phonon dispersion from the rSCAN functional has reduced acoustic-bunching, which reduces the three-phonon scattering phase space \cite{lindsay2013a}. Consequently, the phonon scattering lifetimes obtained from the rSCAN functional are up to an order of magnitude larger than those from the PBEsol functional. The large phonon group velocities, combined with reduced phonon-phonon scattering, result in a factor of two larger $\kappa$ from the rSCAN functional compared to the PBEsol functional for ZnLiAs. }

{\color{black} The above results suggest large variation in predicted $\kappa$ based on employed XC functional. To check if some functionals are more consistent in describing experimentally measured $\kappa$, we looked in the literature for experimentally measured $\kappa$ of the considered materials. However, from 50 materials, we were able to find results for only three materials: (\roml{1}) for BiLiMg in the cubic space group, the experimentally measured $\kappa$ is 5.0 W/m-K for single crystal samples \cite{expmnt1}, compared to 4.6, 5.5 5.0, and 6.2 W/m-K from PBE, LDA, PBEsol, and rSCAN functionals, (\roml{2}) for $\text{CdIn}_2\text{S}_4$ in the cubic spacegroup, the experimentally measured value is 4.5 W/m-K for single crystal samples \cite{expmnt2}, compared to 4.0, 3.2, 4.7, 3.8 from PBE, LDA, PBEsol, and rSCAN functionals, and (\roml{3}) for $\text{AgIn}_5\text{Te}_8$ in the tetragonal spacegroup, the experimentally measured value is 1.08 W/m-K for polycrystalline pellets \cite{expmnt3}, compared to direction-averaged values of 1.7, 1.9, 1.6, 2.0 W/m-K from PBE, LDA, PBEsol, and rSCAN functionals without phonon-boundary scattering.}

\begin{figure}
\begin{center}
\epsfbox{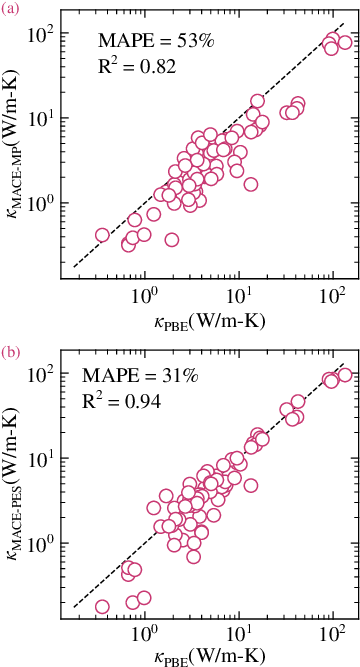}
\end{center}
\caption{Thermal conductivities obtained using foundational ML models trained originally on (a) MP and (b) MatPES datasets \cite{materialsProject, matpes}. The comparison is made against values obtained using the PBE functional as PBE functional was employed in the dataset generation for MP- and MatPES-datasets.}
\label{fig_ML}
\end{figure}

{\bf ML Models:} 
In Fig.~\ref{fig_ML}, we report the comparison of $\kappa$ obtained using MACE foundational ML models based forces with those obtained from DFT-based forces \cite{mace}. The DFT forces are obtained using the VASP package with PBE functional and BTE calculations are carried out using the ALD BTE solver. The PBE functional is chosen for comparison as the underlying ML models were trained on PBE-based DFT calculations. For Fig.~\ref{fig_ML}(a), results are obtained using the MACE ML model where training was performed on the Materials Project dataset (close to equilibrium static configurations) \cite{materialsProject}. For Fig.~\ref{fig_ML}(b), results are obtained using the MACE ML model for which training was performed on the MatPES dataset consisting of DFT/ML-driven molecular dynamics trajectories (out-of-equilibrium configurations) \cite{matpes}. For both cases, we restrict the structures to the DFT relaxed configuration and only obtain forces from the ML models, i.e., structure relaxation is not performed using the ML model. This is done as we noticed that, for certain cases, considered ML models failed drastically in resolving small close-to-equilibrium forces and resulted in a significantly different relaxed structure compared to the DFT counterpart. This limitation of ML models in resolving small close-to-equilibrium forces has been reported in literature, for instance, in Ref.~\cite{srivastava2024}, where we reported small prediction errors from ML forces when trained on thermally perturbed structures compared to larger errors on finite-difference perturbation-based structures.

We find that the $\kappa$ obtained from both ML models follows the correct trend. However, the errors obtained compared to the PBE functional are high: with the MatPES model, the error is somewhat reduced but it stays high at 31\%. Compared to the MP model, which underpredicts  $\kappa$, the MatPES model is unbiased and results in both under- and over-prediction of $\kappa$ compared to DFT-based values.  
{\color{black} The under-prediction of $\kappa$ from the MP model is due to consistent under-prediction of phonon group velocities and over-prediction of lattice anharmonicity for all materials. With MatPES, the performance is improved owing to better prediction of phonon group velocities and lattice anharmonicity (mean absolute percentage error reduced from 20\% for MP model to 14\% for MatPES model for speed of sound and from 21\% for MP model to 10\% for MatPES model for average Gr\"uneissen parameters. }
It is worthwhile to note that these errors obtained from foundational ML models are similar to those reported for direct end-to-end $\kappa$ prediction (without requiring interatomic forces/force constants or BTE solvers). However, the end-to-end ML model performances are highly susceptible to training datasets, while the foundational ML models are more robust/universal.

\section{Discussion}
The thermal transport theory for solids has evolved significantly in the past decade. With the availability of many public-domain DFT packages and BTE solvers, the studies reporting first-principles-based $\kappa$ and their comparison with experiments are now a routine procedure. Many of the theoretical developments in thermal transport theory are motivated by the discrepancy between DFT-based $\kappa$-prediction and experimental measurements. While experimental uncertainties are often accounted for in such comparisons (for instance, through inclusion of boundary or grain-boundary scatterings), the uncertainties associated with the use of employed numerical parameters, such as those discussed above, are rarely reported. 

In this regard, we have shown that there are more than 20\% uncertainties arising from the choice of XC functional itself. The choice of functional is often justified in the literature on the basis of better agreement of lattice constant \cite{jain2015b, xia2020b}, but the agreement of lattice constant may not be the right metric, as presented in Fig.~\ref{fig_XC}. 
This under-prediction of $\kappa$ by the PBE functional is consistent with our previous study on $\text{ThO}_2$, where we found a good agreement of predicted $\kappa$ between LDA and PBEsol functionals and experiments, while the PBE functional resulted in a severe under-prediction \cite{vikrante2025}.  This has also been reported in the literature for other rock salt and zinc-blende binary semiconductors with cubic structures \cite{xia2020b}.

Amongst other parameters, our results show that there is a minimal effect of BTE solver on obtained  $\kappa$ and all solvers result in a good agreement, despite having different implementations of invariance constraints, energy conservation delta functions, etc.  Similarly, we find small differences in $\kappa$ obtained from different DFT packages. For pre-trained foundational ML models, we find that while these models are generic and the accuracy of $\kappa$ obtained based on the forces of these models is improving with the inclusion of more diverse training data (MatPES trained model compared to MP trained model), the accuracies are currently limited (similar to end-to-end ML models \cite{srivastava2023}). Further, our finding of under-prediction of $\kappa$ by the PBE functional, compared to other functionals, may have severe consequences on the foundational ML approaches and high-throughput $\kappa$ datasets, as many of these currently available datasets/ML training are based on the PBE functional-based results. 

{\color{black}Based on our analysis, we find that although rSCAN is a meta-GGA functional, it consistently overestimates the thermal conductivity ($\kappa$), primarily due to its treatment of bond stiffness and its tendency to underestimate anharmonic effects. Among the remaining functionals, we recommend benchmarking the computed phonon frequencies and group velocities against experimental data prior to performing $\kappa$ calculations. In cases where experimental validation is not feasible, the use of the PBEsol functional is advised, as the resulting $\kappa$ values typically lie within the bounds defined by PBE and LDA functionals. However, caution should be exercised when interpreting these results, especially when developing new physical models, to ensure that the predicted $\kappa$ values do not have accidental agreement with experimental observations.}

{\color{black}Finally, we note that the uncertainties arising from the use of different force constant extraction approaches (finite-difference vs density functional perturbation theory for harmonic force constants, finite-difference vs compressing sensing, etc, for anharmonic force constants, effect of step-size, etc) and thermal transport physics (multi-channel thermal transport, phonon renormalization, higher-order scattering, etc) are not reported in this study. Several of such effects are reported to be important for low-dimensional and high-$\kappa$ materials in the literature \cite{jain2020, han2023, zhou2023}. }

\section{Conclusions}
In summary, we have carried out a computational benchmarking of uncertainties in phonon thermal conductivity calculation of semiconductors using the BTE approach. We considered uncertainties originating from the use of different BTE solvers, DFT packages, XC functionals, and foundational ML models. Our results suggest that while the variations in thermal conductivity from BTE solvers and DFT packages are small with MAPE $<$ 10\%, the use of different XC functionals results in a large variation ($>$ 20\% MAPE). 
The use of foundational ML models results in the right trend of thermal conductivity, but they result in additional $>$ 30\% MAPE compared to the PBE functional data on which the models were trained.

\section{Acknowledgement}
The authors acknowledge the financial support from the Core Research Grant, Science \& Engineering Research Board, India (Grant Number: CRG/2021/000010) and Nano
Mission, Government of India (Grant No.: DST/NM/NS/2020/340).  The calculations are carried out on the SpaceTime-II supercomputing facility of IIT Bombay.

\section{Data Availability}
The data required to reproduce the findings are available from the corresponding author on reasonable request. 

\bibliography{references}

\begin{thebibliography}{40}
\bibitem [{Ta_()}]{Ta_note}%
  \BibitemOpen
  \href@noop {} {}\bibinfo {note} {Due to non-availability of other Ta-based materials in the considered dataset, we are unable to ascertain if this outlier from rSCAN is specific to Ta.}\BibitemShut {Stop}%
\end{thebibliography}

\clearpage
\onecolumngrid
\appendix
\label{table_data}

\section{The lattice thermal conductivity (in W/m-K) of the considered materials as obtained using different settings. For materials in the trigonal and tetragonal space groups, the properties from two non-equal directions are reported as different datapoints. For QE, all calculations are performed using the PBE functional by employing three different BTE solvers. For VASP, all calculations are performed using the ALD BTE solver by employing different XC functionals. For MACE ML models, the calculations are performed using the ALD BTE solver.}
\begin{longtable}{| r | M{1.8cm} | M{1.cm}  M{1.5cm}  M{1.5cm} || M{1cm} M{1cm} M{1.2cm} M{1cm} || M{1cm} M{1cm} |}\hline
Stoichiometry & SpaceGroup & \multicolumn{3}{|c|}{QE} & \multicolumn{4}{|c|}{VASP} & \multicolumn{2}{|c|}{MACE-ML} \\
\cline{3-5}
\cline{6-9}
\cline{10-11}
& Number & ALD & ShengBTE & Phono3py & PBE & LDA & PBEsol & rSCAN & MP & PES \\
\hline
 Al$_2$CdS$_4$  &   82 &    5.81 &    5.98 &    5.95 &   4.81 &   5.84 &   5.50  &   6.81 &   4.17 &   6.16 \\
  Al$_2$CdS$_4$  &   82 &    4.39 &    4.50 &    4.48 &   3.83 &   4.80 &   4.60  &   5.32 &   2.83 &   3.72 \\
  CdGa$_2$S$_4$  &   82 &    5.64 &    5.69 &    5.67 &   6.77 &   5.49 &   5.63  &   5.49 &   5.21 &   4.17 \\
  CdGa$_2$S$_4$  &   82 &    5.62 &    5.65 &    5.65 &   5.98 &   5.45 &   5.14  &   4.60 &   4.33 &   3.23 \\
 CdGa$_2$Se$_4$  &   82 &    3.78 &    3.80 &    3.79 &   3.53 &   3.34 &   3.22  &   3.34 &   3.86 &   3.43 \\
 CdGa$_2$Se$_4$  &   82 &    3.35 &    3.36 &    3.35 &   3.20 &   2.88 &   2.89  &   3.11 &   2.76 &   3.32 \\
 CdGa$_2$Te$_4$  &   82 &    3.26 &    3.27 &    3.26 &   3.81 &   3.71 &   3.65  &   4.50 &   1.07 &   2.06 \\
 CdGa$_2$Te$_4$  &   82 &    2.64 &    2.65 &    2.65 &   3.04 &   3.00 &   2.92  &   3.62 &   0.93 &   1.66 \\
 Al$_2$Se$_4$Zn  &   82 &    7.10 &    7.13 &    7.12 &   7.27 &   7.38 &   5.21  &  11.19 &   5.58 &   4.80 \\
 Al$_2$Se$_4$Zn  &   82 &    6.34 &    6.38 &    6.36 &   6.57 &   6.60 &   4.53  &   8.81 &   5.38 &   4.57 \\
 Al$_2$BaTe$_4$  &   97 &    2.43 &    2.46 &    2.45 &   2.31 &   3.48 &   3.04  &   3.07 &   1.73 &   1.89 \\
 Al$_2$BaTe$_4$  &   97 &    0.62 &    0.65 &    0.63 &   0.67 &   1.18 &   0.96  &   0.83 &   0.33 &   0.43 \\
 CdIn$_2$Se$_4$  &  111 &    4.88 &    4.90 &    4.87 &   4.95 &   5.03 &   5.15  &   5.39 &   3.83 &   4.01 \\
 CdIn$_2$Se$_4$  &  111 &    3.24 &    3.26 &    3.25 &   3.65 &   3.72 &   3.70  &   4.07 &   2.45 &   3.67 \\
 AgIn$_5$Se$_8$  &  111 &    2.45 &    2.47 &    2.45 &   2.10 &   2.02 &   1.79  &   2.24 &   2.34 &   1.25 \\
 AgIn$_5$Se$_8$  &  111 &    0.88 &    0.88 &    0.88 &   0.74 &   0.73 &   0.56  &   0.70 &   0.39 &   0.20 \\
 AgIn$_5$Te$_8$  &  111 &    2.11 &    2.13 &    2.12 &   2.05 &   2.29 &   1.98  &   2.40 &   0.99 &   0.95 \\
 AgIn$_5$Te$_8$  &  111 &    1.13 &    1.14 &    1.13 &   0.98 &   1.19 &   0.94  &   1.10 &   0.42 &   0.23 \\
 CuGa$_5$Se$_8$  &  111 &    6.06 &    6.06 &    6.06 &   5.38 &   5.64 &   5.57  &   5.85 &   4.13 &   2.12 \\
 CuGa$_5$Se$_8$  &  111 &    5.28 &    5.24 &    5.30 &   3.33 &   3.34 &   3.43  &   3.41 &   2.26 &   1.00 \\
 Mg$_4$SeTe$_3$  &  115 &    9.72 &    9.74 &    9.82 &   9.53 &   6.79 &   6.88  &  12.05 &   6.98 &   8.75 \\
 Mg$_4$SeTe$_3$  &  115 &    7.05 &    7.12 &    7.20 &   7.12 &   4.88 &   5.17  &   8.68 &   4.33 &   5.25 \\
      AlGaN$_2$  &  115 &  138.61 &  145.86 &  145.16 & 133.27 & 173.92 & 157.61  & 167.03 &  76.64 &  94.51 \\
      AlGaN$_2$  &  115 &   98.30 &  103.27 &  102.90 &  95.38 & 128.06 & 111.02  & 119.89 &  64.98 &  79.29 \\
      AlGaP$_2$  &  115 &   55.83 &   57.49 &   57.29 &  42.17 &  46.43 &  45.36  &  51.00 &  14.75 &  46.11 \\
      AlGaP$_2$  &  115 &   44.23 &   45.02 &   44.97 &  32.01 &  36.42 &  35.29  &  37.28 &  11.47 &  37.12 \\
     As$_2$GaIn  &  115 &   12.00 &   12.08 &   12.08 &  16.67 &  19.25 &  17.65  &  20.72 &   8.10 &  17.23 \\
     As$_2$GaIn  &  115 &   11.30 &   11.53 &   11.73 &  15.03 &  17.49 &  16.15  &  17.59 &   7.21 &  14.31 \\
      AsIn$_2$P  &  115 &   45.86 &   46.11 &   46.03 &  41.12 &  44.79 &  41.35  &  47.25 &  12.89 &  30.33 \\
      AsIn$_2$P  &  115 &   39.94 &   40.54 &   40.41 &  36.78 &  39.22 &  37.27  &  41.67 &  11.48 &  28.53 \\
      Mg$_2$SSe  &  115 &   18.15 &   18.22 &   18.28 &  17.65 &  15.78 &  15.59  &  20.79 &   8.92 &  16.59 \\
      Mg$_2$SSe  &  115 &   13.88 &   14.01 &   13.99 &  13.42 &  12.15 &  11.92  &  15.79 &   6.80 &  13.36 \\
 In$_2$MgTe$_4$  &  121 &    4.56 &    4.58 &    4.57 &   3.98 &   4.35 &   3.90  &   4.85 &   2.06 &   1.36 \\
 In$_2$MgTe$_4$  &  121 &    3.77 &    3.82 &    3.79 &   3.29 &   2.85 &   2.48  &   3.94 &   1.32 &   0.69 \\
     GeMg$_4$Si  &  123 &    5.73 &    5.73 &    5.72 &   5.79 &   7.00 &   6.31  &   7.24 &   2.17 &   5.52 \\
     GeMg$_4$Si  &  123 &    5.18 &    5.17 &    5.17 &   5.05 &   5.82 &   5.38  &   6.53 &   1.92 &   4.97 \\
     Na$_4$SeTe  &  123 &    2.19 &    2.19 &    2.18 &   2.12 &   2.83 &   2.34  &   2.71 &   1.51 &   1.91 \\
     Na$_4$SeTe  &  123 &    1.89 &    1.87 &    1.86 &   1.80 &   2.37 &   1.98  &   2.24 &   1.23 &   1.58 \\
As$_2$Be$_2$Li$_2$  &  129 &    9.46 &    9.53 &    9.61 &   9.48 &  12.58 &  12.09  &   9.92 &   2.38 &   9.96 \\
As$_2$Be$_2$Li$_2$  &  129 &    3.22 &    3.14 &    3.25 &   2.90 &   4.11 &   3.61  &   3.15 &   1.09 &   3.93 \\
Li$_2$Na$_2$S$_2$  &  129 &    3.65 &    3.66 &    3.70 &   3.58 &   5.37 &   4.40  &   4.72 &   3.18 &   2.94 \\
Li$_2$Na$_2$S$_2$  &  129 &    2.75 &    2.75 &    2.74 &   2.68 &   4.21 &   3.57  &   3.91 &   2.75 &   2.70 \\
  BaCd$_2$P$_2$  &  164 &    4.47 &    4.47 &    4.48 &   4.31 &   7.33 &   6.34  &   5.99 &   2.03 &   4.42 \\
  BaCd$_2$P$_2$  &  164 &    3.53 &    3.53 &    3.55 &   3.21 &   4.35 &   4.16  &   4.73 &   1.39 &   3.18 \\
  BaMg$_2$P$_2$  &  164 &    5.60 &    5.58 &    5.58 &   5.41 &   6.38 &   5.84  &   6.40 &   4.58 &   5.74 \\
  BaMg$_2$P$_2$  &  164 &    3.68 &    3.65 &    3.67 &   3.79 &   4.05 &   3.68  &   4.49 &   2.53 &   3.38 \\
 BaMg$_2$Sb$_2$  &  164 &    6.50 &    6.55 &    6.57 &   5.72 &   8.13 &   6.95  &   6.69 &   2.71 &   6.16 \\
 BaMg$_2$Sb$_2$  &  164 &    6.41 &    6.44 &    6.48 &   4.92 &   6.94 &   6.30  &   6.32 &   2.05 &   5.04 \\
     AgScSe$_2$  &  164 &    0.73 &    0.73 &    0.73 &   0.78 &   1.27 &   1.14  &   1.13 &   0.63 &   0.49 \\
     AgScSe$_2$  &  164 &    0.36 &    0.37 &    0.36 &   0.35 &   0.88 &   0.69  &   0.53 &   0.42 &   0.18 \\
Be$_2$Li$_2$Sb$_2$  &  186 &   13.86 &   14.54 &   15.43 &  10.30 &  12.98 &  13.28  &   9.80 &   3.94 &   8.45 \\
Be$_2$Li$_2$Sb$_2$  &  186 &   11.49 &   11.55 &   11.62 &   8.94 &  10.91 &  11.31  &   8.84 &   3.03 &   5.83 \\
         AsBaLi  &  187 &    3.15 &    3.16 &    3.18 &   3.59 &   4.36 &   3.92  &   4.03 &   1.90 &   3.33 \\
         AsBaLi  &  187 &    3.12 &    3.12 &    3.13 &   2.95 &   3.96 &   3.25  &   3.28 &   1.60 &   2.75 \\
As$_4$In$_4$Sr$_2$  &  194 &    3.21 &    3.20 &    3.18 &   3.00 &   3.86 &   3.52  &   4.88 &   3.37 &   4.94 \\
As$_4$In$_4$Sr$_2$  &  194 &    1.95 &    1.98 &    2.34 &   1.68 &   2.23 &   2.13  &   3.21 &   1.35 &   3.58 \\
Li$_2$S$_4$Ta$_2$  &  194 &   12.50 &   12.56 &   12.50 &  13.34 &  16.45 &  16.17  &  15.18 &   1.65 &   4.74 \\
Li$_2$S$_4$Ta$_2$  &  194 &    1.67 &    1.72 &    1.82 &   1.93 &   2.44 &   2.33  &   1.31 &   0.37 &   1.61 \\
   Mg$_4$Te$_8$  &  205 &    3.23 &    3.24 &    3.40 &   3.48 &   3.71 &   3.58  &   4.30 &   1.36 &   3.20 \\
  AlGa$_3$N$_4$  &  215 &  106.57 &  107.97 &  107.83 &  98.77 & 126.26 & 115.68  & 116.67 &  84.40 &  84.38 \\
  Al$_3$GaN$_4$  &  215 &   95.69 &   95.63 &   95.16 &  90.01 & 113.34 & 107.69  & 108.18 &  74.36 &  84.40 \\
Li$_8$Mg$_4$Si$_4$  &  215 &    1.05 &    1.05 &    1.05 &   1.24 &   1.62 &   1.53  &   1.68 &   0.73 &   2.59 \\
  Al$_5$CuS$_8$  &  216 &    4.18 &    4.20 &    4.17 &   3.67 &   4.52 &   5.61  &   3.80 &   5.83 &   3.44 \\
         AsCdLi  &  216 &    4.17 &    4.63 &    4.72 &   4.06 &   6.04 &   5.49  &   4.77 &   5.08 &   3.55 \\
         BiLiMg  &  216 &    4.35 &    4.42 &    4.43 &   4.61 &   5.51 &   5.04  &   6.21 &   2.12 &   6.96 \\
          LiMgN  &  216 &   13.87 &   13.40 &   13.85 &  15.67 &  21.45 &  17.43  &  20.96 &  15.75 &  18.86 \\
          LiMgP  &  216 &    8.69 &    8.76 &    8.79 &   8.37 &   9.92 &   9.34  &  10.71 &   5.82 &   9.62 \\
         AsLiZn  &  216 &    4.62 &    4.67 &    4.62 &   4.19 &   6.63 &   6.38  &  13.66 &   2.86 &   6.17 \\
          LiNZn  &  216 &   13.31 &   13.39 &   13.36 &  14.09 &  16.90 &  16.10  &  12.79 &  10.97 &  14.87 \\
          LiPZn  &  216 &    6.73 &    6.96 &    6.86 &   6.81 &  10.29 &   8.58  &   7.33 &   4.20 &   8.21 \\
      Ca$_3$NSb  &  221 &    5.46 &    5.46 &    5.46 &   4.96 &   5.22 &   5.46  &   4.16 &   6.35 &   5.15 \\
     BiLi$_2$Na  &  225 &    1.44 &    1.43 &    1.43 &   1.47 &   1.97 &   1.70  &   1.67 &   1.25 &   1.56 \\
Cd$_2$In$_4$S$_8$  &  227 &    3.12 &    3.11 &    3.10 &   4.00 &   3.24 &   4.72  &   3.77 &   2.27 &   1.33 \\
Cd$_2$Sc$_4$Se$_8$  &  227 &    2.60 &    2.61 &    2.60 &   2.60 &   1.83 &   2.22  &   2.90 &   3.34 &   3.13 \\
Cd$_2$S$_8$Y$_4$  &  227 &    3.01 &    3.01 &    3.00 &   2.97 &   3.21 &   3.75  &   4.55 &   1.88 &   3.21 \\
Cd$_2$Se$_8$Y$_4$  &  227 &    2.00 &    2.01 &    2.00 &   2.05 &   1.73 &   2.01  &   2.51 &   1.65 &   2.58 \\
Mg$_2$Se$_8$Y$_4$  &  227 &    3.08 &    3.07 &    3.07 &   3.24 &   3.13 &   2.72  &   4.76 &   4.34 &   3.39 \\
B$_3$Ca$_4$LiN$_6$  &  229 &    2.54 &    2.57 &    2.61 &   3.40 &   3.80 &   3.98  &   4.07 &   1.67 &   2.73 \\
Cd$_2$In$_4$S$_8$  &  227 &    3.12 &    3.11 &    3.10 &   4.00 &   3.24 &   4.72  &   3.77 &   2.27 &   1.33 \\
Cd$_2$Sc$_4$Se$_8$  &  227 &    2.60 &    2.61 &    2.60 &   2.60 &   1.83 &   2.22  &   2.90 &   3.34 &   3.13 \\
Cd$_2$S$_8$Y$_4$  &  227 &    3.01 &    3.01 &    3.00 &   2.97 &   3.21 &   3.75  &   4.55 &   1.88 &   3.21 \\
Cd$_2$Se$_8$Y$_4$  &  227 &    2.00 &    2.01 &    2.00 &   2.05 &   1.73 &   2.01  &   2.51 &   1.65 &   2.58 \\
Mg$_2$Se$_8$Y$_4$  &  227 &    3.08 &    3.07 &    3.07 &   3.24 &   3.13 &   2.72  &   4.76 &   4.34 &   3.39 \\
B$_3$Ca$_4$LiN$_6$  &  229 &    2.54 &    2.57 &    2.61 &   3.40 &   3.80 &   3.98  &   4.07 &   1.67 &   2.73 \\
\hline
\end{longtable}   

\end{document}